# Global Fishing Watch: Bringing Transparency to Global Commercial Fisheries

Wessley Merten

Oceana

Washington, D.C.

wmerten@oceana.org

Adam Reyer

Oceana

Washington, D.C.

areyer@oceana.org

Jackie Savitz

Oceana

Washington, D.C.

jsavitz@oceana.org

John Amos

SkyTruth

Shepherdstown, WV

john@skytruth.org

Paul Woods

SkyTruth

Shepherdstown, WV

paul@skytruth.org

Brian Sullivan

Google

Mountain View, CA

bsullivan@google.com

**ABSTRACT**

Across all major industrial fishing sectors, overfishing due to overcapacity and lack of compliance in fishery governance has led to a decline in biomass of many global fish stocks. Overfishing threatens ocean biodiversity, global food security, and the livelihoods of law abiding fishermen. To address this issue, Global Fishing Watch (GFW) was created to bring transparency to global fisheries using computer science and big data analytics. A product of a partnership between Oceana, SkyTruth and Google, GFW uses the Automatic Identification System, or AIS, to analyze the movement of vessels at sea. AIS provides vessel location data, and GFW uses this information to track global vessel movement and apply algorithms to classify vessel behavior as "fishing" or "non-fishing" activity. Now publicly available, anyone with an internet connection can monitor when and where trackable commercial fishing appears to be occurring around the world. Hundreds of millions of people around the world depend on our ocean for their livelihoods, and many more rely on it for food. Collectively, the various applications of GFW will help reduce overfishing and illegal fishing, restore the ocean's abundance, and ensure sustainability through better monitoring and governance of our marine resources.

## 1. INTRODUCTION

The ocean is vast and largely still unexplored and undocumented[1]. In terms of governance, our marine resources remain grossly unmanaged and inadequately protected[2]. In recent years, however, there have been attempts to protect marine ecosystems and biodiversity by creating large marine protected areas and developing a framework for a global treaty to better manage those areas and the resources within them on the high seas[3,4]. Historically, these efforts have been constrained by humanity's lack of capacity to effectively monitor and enforce non-compliance in such vast areas, but a fast-growing publically accessible source of ship-tracking data known as AIS has potential to change that. Vessels broadcast and receive AIS signals to avoid colliding with each other. Satellite-based receivers can also pick up these signals, and AIS is increasingly being viewed not only as an important maritime safety system, but as an equally important tool for monitoring and enforcing our oceans to diminish illegal activity. One major sector that could benefit from this new application of AIS data is fisheries.

In 2009, a report estimated the annual economic and biological impact of Illegal, Unreported and Unregulated fishing (IUU) on global fisheries to be a $10-23.5 billion loss in revenue from an illegal harvest of 11-26 million tonnes of fish[3]. The report also found that developing countries and Small Island Developing States (SIDS) were more prone to IUU activity due to an inability to govern their sovereign waters. For example, estimated catches in West African and Western Central Pacific Exclusive Economic Zones (EEZs), were 37% and 34%, higher than what was actually being reported due to lack of enforcement of foreign fishing fleet license and

reporting requirements. These estimates not only represent a major loss of information necessary to conduct proper scientific assessments of the health of fish stocks in the region, but also an economic and biological loss for these nations. Most of the catch is landed elsewhere and in contravention of sustainable fisheries practices designed to ensure future conservation of economically important fisheries[4].

The solution to resolving the problem of IUU fishing is multi-faceted, but the single most important component begins with fisheries monitoring. Global Fishing Watch, a product of a partnership between Oceana, SkyTruth, and Google, is making the impossible, possible, by providing the first open-source global near-real time record of apparent fishing effort. Citizens can use it to see for themselves whether their fisheries are being effectively managed, and hold their leaders accountable for long-term fishery sustainability. Seafood suppliers can keep tabs on the vessels from which they buy fish. Journalists and the public can act as watchdogs to improve the sustainable management of global fisheries. Responsible fishermen can use Global Fishing Watch to show they are adhering to the law and researchers can access a multi-year record of all apparent fishing activity to conduct applied scientific research and thus policy goals for marine conservation. Hundreds of millions of people around the world depend on our ocean for their livelihoods, and many more rely on it for food. Collectively, the various applications of Global Fishing Watch will help reduce overfishing and illegal fishing and help restore the ocean's abundance and ensure sustainability through better monitoring and governance of our marine resources.

**2. METHODS**
Global Fishing Watch harnesses data collected from land and satellite-based receivers. AIS signals, emitted as frequently as several times per minute, include the position, speed, heading, vessel identity, and other information. Our AIS database spans January 1, 2012, until the present and contains more than 20 billion AIS messages processed from more than 200,000 ocean going vessels. To examine fishing vessels contained within this database, Global Fishing Watch developed a behavioral classification model to identify when and where vessels appear to be engaged in fishing activity. The output of the classification model is apparent fishing activity, measured in the number of hours each vessel spent engaged in apparent fishing activity, displayed on an interactive map. The fishing vessels displayed on the interactive map have been classified into three categories: (1) known fishing vessels; (2) likely fishing vessels; (3) suspected fishing vessels. Known fishing vessels have been matched by Marine Maritime Service Identity (MMSI) number to unique vessel identifiers such as International Maritime Organization (IMO), Common List of Authorized Vessel (CLAV) number, International Radio Callsign (IRCS), ship name, and gear type from major vessel registries (e.g., CLAV list, Regional Fisheries Management Organization lists, etc.). Likely fishing vessels are vessels that self-identify as fishing vessels in their AIS messages. Suspected fishing vessels are labeled as such based on repeated detection of apparent fishing activity.

**3. RESULTS**
The Beta version of Global Fishing Watch contains 20 billion processed AIS messages and initially includes 35,000 fishing vessels. The first study to use these data was recently published in the journal Science (see McCauley et al. 2016), showcasing the applicability of AIS as a credible data source for conducting applied scientific research. The authors compared apparent fishing effort classified from AIS to fisheries observer data obtained from the Western and Central Pacific Fisheries Commission and found significant positive correlation between the datasets. Based on these results, AIS was then used to monitor the shift in fishing effort before and after commercial fishing was banned in the Phoenix Islands Protected Area (PIPA) in the central Pacific Ocean on January 1, 2015. The results showed that vessels complied with the closure, maintained their use of AIS, and shifted their fishing effort outside of PIPA. Interestingly, while McCauley et al. (2016) highlighted the applicability of using AIS to monitor Marine Protected Areas (MPAs), Global Fishing Watch's dataset can also be used for enforcement of laws and regulations governing MPAs and fisheries. In May 2015, AIS data processed by Global Fishing Watch was used in an illegal fishing event in PIPA when a Marshall Islands flagged purse seiner, the Marshalls 203, was observed to have fished in PIPA after the ban came into effect; the use Global Fishing Watch data helped result in a $1 million fine being paid by the Central Pacific Fishing Company, the owners of the Marshalls 203. The company also made a special "goodwill arrangement" wit hteh Government of Kiribati, agreeing to pay an additional $1 million in the form of a grant. This event not only highlights the use of Global Fishing Watch and AIS data as a fisheries monitoring tool, but also as an aid to enforcement.

**4.CONCLUSION**
Originally designed as a collision avoidance system to ensure the safety of vessels and their crews at sea, it is widely recognized now that the availability, resolution, and nature of AIS data can play an equally important role in mitigating overfishing and IUU fishing, to safeguard the health and conservation of global fish stocks for generations to come[6]. Global Fishing Watch is the first open source and freely available data source and technology platform to visualize apparent fishing effort using AIS data. This data-driven platform will allow anyone with an internet connection to view the movements of global fishing fleets, will enable myriad third-party applications and research initiatives, and will spark questions about how data science can be used to address fishery policy needs. According to World Bank, the world will need 30-49 million metric tons more seafood, roughly doubling our edible seafood supply, to meet the demand by 2030[8]. Sustainable management of wild capture fisheries is inextricably tied to meeting this goal. Global Fishing Watch will provide a vital tool to verify that fishing fleets are complying with global conservation and management measures, to ultimately save our oceans and feed the world.

**5.ACKNOWLEDGEMENTS**
Global Fishing Watch acknowledges the support of all our funders.

**6.REFERENCES**